\documentclass[galaxies,letter,accept,moreauthors,pdftex,10pt,a4paper,usenatbib]{mdpi} 
\firstpage{1} 
\articlenumber{x}
\doinum{10.3390/------}
\pubvolume{xx}
\pubyear{2017}
\copyrightyear{2017}
\externaleditor{Academic Editor: name}
\history{Received: date; Accepted: date; Published: date}

\usepackage{epsfig}
\usepackage{amsmath}
\usepackage{amssymb}
\usepackage{natbib}
\usepackage{multirow}
\usepackage{graphicx} %\graphicspath{{graphics/}}
\usepackage{color}
\usepackage{epstopdf}
\usepackage{aas_macros}

\Title{Distribution and Evolution of Metals in the Magneticum simulations}

\Author{Klaus Dolag$^{1,2}$, Emilio Mevius$^{1}$, and Rhea-Silvia Remus $^{1}$}

\AuthorNames{Klaus Dlag and Emilio Mevius \& Rhea-Silvia Remus}

\address{%
$^{1}$ \quad Universit\"ats-Sternwarte M\"unchen, Scheinerstr.\ 1, D-81679 M\"unchen, Germany\\
$^{2}$ \quad Max Planck Institut for Astrophysics, D-85748 Garching, Germany}

\corres{dolag@usm.lmu.de}

\abstract{Metals are ideal tracers of the baryonic cycle within
  halos. Their composition is a fossil record connecting the evolution
  of the various stellar components of galaxies to the interaction
  with the environment by in- and outflows.  The {\it Magneticum}
  simulations allow to study halos across a large range of masses and
  environments, from massive galaxy clusters containing hundreds of
  galaxies down to isolated field galaxies. They include a detailed
  treatment of the chemo-energetic feedback from the stellar component
  and its evolution as well as feedback from the evolution of
  supermassive black holes. Following the detailed evolution of
  various metal species and their relative composition due to
  continuing enrichment of the IGM and ICM by SNIa, SNII and AGB winds
  of the evolving stellar population reveals the complex interplay of
  local star formation processes, mixing, global baryonic flows,
  secular galactic evolution and environmental processes. We present
  results from the {\it Magneticum} simulations on the chemical
  properties of simulated galaxies and galaxy clusters, carefully comparing them to
  observations. We show that the simulations already reach a very high
  level of realism within their complex descriptions of the
  chemo-energetic feedback, successfully reproducing a large number of
  observed properties and scaling relations. Our simulated galaxies
  clearly indicate that there are no strong secondary parameters (like
  star formation rates at fixed redshift) driving the scatter in these
  scaling relations. The remaining differences clearly point to
  detailed physical processes which have to be included into future
  simulations.}

\keyword{Galaxy clusters -- intra cluster medium -- galaxies -- stellar population -- numerical simulation}

\begin{document}

%%%%%%%%%%%%%%%%%%%%%%%%%%%%%%%%%%%%%%%%%%
\section{Introduction}

In cosmological hydrodynamical simulations, star formation is
typically treated based on a sub-grid model. Star particles are formed
when the gas locally exceeds a certain density threshold, when it is
Jeans unstable, and the local gas flow converges at a gas
particle. Each star particle represents a simple stellar population
which is characterized by an initial mass function (IMF). 

In addition, each star particle emits a feedback that is mimicking the
winds which are launched by stars in reality by averaging over the
whole stellar population included in each simulated stellar particle.
In most models, the considered winds compose of three main stellar
sources: Supernovae Type II (SNII), Supernovae Type Ia (SNIa), and
asymptotic giant branch stars (AGBs).   These three sources are
considered to be the most important drivers of observed stellar
outflows and thus the main contributers to metal enrichment of the
interstellar medium:

Stars with masses larger than 8 $M_\odot$ are believed to end their
lives through a so called core collapse (SNII), releasing typically an
energy of $10^{51}$~erg per supernova together with their chemical
imprint into the surrounding gas. These supernovae usually occur
shortly after the formation of the simulated star particle, as the
most massive stars only live for a short time and thus this part of
the stellar population of the star particle is emitting first.  SNIa
on the other hand are believed to arise from thermonuclear explosions
of white dwarfs within binary stellar systems. These supernovae are
delayed relatively to the SNII and occur later in a star particle's
live, as the white dwarf in the according binary system has to first
be formed and second needs to  accrete matter from its companion until
it reaches the mass threshold for the onset of thermonuclear
burning. Therefore, a detailed modelling of the evolution of the
stellar population has to be included in the simulations to follow the
chemo-energetic imprint of the SNIa on the surrounding gas.  AGB stars
contribute emitting strong winds, exhibiting strong mass losses during
their life and are importantly contributing to the nucleosynthesis of
heavy elements. 

To include the effects of these sources into the sub-grid models of
simulated stellar particles raises the need to integrate a set of
complicated equations describing the evolution of a simple stellar
population. Such set of integral equations then allows to compute at
each time the rate at which the current AGB stars pollute their environment
by stellar winds and the rate at which the SNIa and SNII are exploding.
Such they allow to properly treat the chemo-energetic imprint on the
surrounding inter-galactic medium (IGM) and the inter-cluster medium (ICM).
In the following we
will give a short, schematic description of such calculations based on
the description presented in Dolag et al., (2017)~\cite{Dolag2017} (for
a more detailed review, see Matteucci 2003~\cite{Matteucci2003} and
Borgani et al. 2008~\cite{Borgani2008}, and references
therein). Then we will introduce the cosmological hydrodynamical
simulation set used in this work, and present a comparison of the
metal enrichment in galaxies and clusters of galaxies caused by this
model feedback with observations. Finally, we will summarize and
discuss the results of this comparison in the light of model
improvements needed for the future.

%%%%%%%%%%%%%%%%%%%%%%%%%%%%%%%%%%%%
\section{Chemical Enrichment}

To describe the continuous enrichment of the IGM and ICM through winds
from SNIa, SNII and AGB stars in the evolving stellar population of
each star particle, several ingredients are needed. In the following
we will shortly present the ingredients used in many state-of-the-art
cosmological simulations of galaxy formation.

\subsection{IMF}
One of the most important quantities in models of chemical evolution
is the IMF. It directly determines the
relative ratio between SNII and SNIa, and therefore the relative
abundance of $\alpha$-elements and Fe-peak elements. The shape of the
IMF also determines the ratio between low-mass long-living and massive
short-living stars. This ratio directly affects the amount of energy
released by the SNe as well as the present luminosity of galaxies
which is dominated by low mass stars and the (metal) mass-locking in
the stellar phase.

The IMF $\phi(m)$ describes the number of stars of a given mass per
unit logarithmic mass interval. Historically, a commonly used form is
the Salpeter IMF (Salpeter 1955~\cite{1955ApJ...121..161S}) which
follows a single power-law with an index of $x=1.35$. However,
different expressions of the IMF have been proposed more recently in
order to model a flattening in the low-mass regime of the stellar mass
function that is currently favoured by a number of observations. Among
the newer, often used models is the Chabrier IMF (Chabrier
2003~\cite{2003PASP..115..763C}), which has a continuously changing
slope and is more top heavy than the Salpeter IMF:
\begin{equation}
\phi(m)\, \propto \, \left\{\begin{array}{ll}
m^{-1.3} & m > 1\, {\rm M}_\odot\\
\vspace{-0.2cm}&\\
e^{\frac{-\left(\log(m) -
        \log(m_c)\right)^2}{2\, \sigma^2}} & m \le 1\, {\rm
  M}_\odot\\
\end{array}\right.
\label{Chabrier}
\end{equation}
However, the question of whether there is a global IMF or if the IMF
is changing with galaxy mass, morphology or cosmological time, and
which IMF has to be chosen is still an unsolved problem and a matter
of heavy debate.

\subsection{Lifetime functions}
To model the evolution of a simple stellar population, a detailed
knowledge of the lifetimes of stars with different masses is required.
Different choices for the mass-dependence of the lifetime function
have been proposed in the literature (e.g. Padovani and Matteucci
(1993)~\cite{1993ApJ...416...26P}, Maeder and Meynet
(1989)~\cite{1989A&A...210..155M}, Chiappini et al.,
(1997)~\cite{1997ApJ...477..765C}), where the latest one reads:
\begin{equation}
\tau(m)=\left\{\begin{array}{ll}
10^{ -0.6545 \log m + 1} & m \le 1.3~{\rm M}_\odot\\
\vspace{-0.2cm}&\\
10^{ -3.7 \log m + 1.351} & 1.3 < m \le 3~{\rm M}_\odot\\
\vspace{-0.2cm}&\\
10^{ -2.51 \log m + 0.77} & 3 < m \le 7~{\rm M}_\odot\\
\vspace{-0.2cm}&\\
10^{ -1.78 \log m + 0.17} & 7 < m \le 15~{\rm M}_\odot\\
\vspace{-0.2cm}&\\
10^{ -0.86 \log m - 0.94} & 15 < m \le 53~{\rm M}_\odot\\
\vspace{-0.3cm}&\\
1.2 \times m^{-1.85}+ 0.003 & {\rm otherwise.} 
\end{array} \right.
\end{equation}

%%%%%%%%%%%%%%%%%%%%%%%%%%%%%
\subsection{Stellar Yields}
The ejected masses of the different metal species $i$ produced by a
star of mass $m$ are called stellar yields {\it
  p$_{Z_{i}}$(m,Z)}. These yields depend on the metallicity
$Z$ with which the star originally formed, and on the type of outflow
ejected from the star.  Therefore, detailed predictions for the main
three sources of enrichment are needed, namely for the mass loss of
AGB stars, of SNII, and of SNIa. Up to date, such predictions still
suffer from significant uncertainties, mainly due to the still poorly
understood mass loss through stellar winds in stellar evolution
models, which depends on multiple additional physical processes.

For the mass loss through AGB stars, the most recent predictions can
be found in Karakas (2007)~\cite{2007PASA...24..103K}.  Predictions
for the mass loss from massive stars driving SNII are presented by
Nomoto, Kobayashi \& Tominaga (2013)~\cite{2013ARA&A..51..457N}.  The
most complete table for SNIa up to date is presented by Thielmann
(2003)~\cite{2003fthp.conf..331T}.

\subsection{Modelling the Enrichment Process}
As summarized by Dolag (2017)~\cite{Dolag2017}, the assumption of a
generic star formation history represented by an arbitrary function
 of time $\psi(t)$ allows to compute the rates for the different
contributions in form of a set of integral equations as shown
in the following (for more details, see also Matteucci 2003~\cite{Matteucci2003}, Borgani et
al. 2008~\cite{Borgani2008}, and references therein).
This formalism can be individually applied to the large number of particles
representing the continuous star-formation process within cosmological
simulations. Every star particle here represents a stellar population
born in a single burst. The combination of all the stellar component within
the simulated galaxies results then in a model which describes
the legacy of the detailed star-formation history of any simulated galaxy.

\subsubsection{Type Ia supernovae}
SNIa occur in binary systems with masses in a mass range of 0.8--8
M$_\odot$. Let $m_{\mathrm B}$ be the total mass of the binary system
and $m_2$ the mass of the secondary companion. We can now use $f(\mu)$
as the distribution function of binary systems with
$\mu=m_2/m_{\mathrm B}$ and define $A$ as the fraction of stars in
binary systems that are progenitors of SNIa. Therefore, $A$ has to be
given or obtained by a model. Constructing such detailed models for
SNIa progenitors is particular difficult, see for example Greggio \& Renzini
(1983)~\cite{1983A&A...118..217G} or Greggio (2005)~\cite{2005A&A...441.1055G}. 
Based on such kind of models, typical value for $A$ are inferred to be
in the range of $0.05$ and $0.1$, based on comparisons of chemical enrichment
models with observed iron metallicities within galaxy clusters (e.g. Matteucci
\& Gibson 1995~\cite{1995A&A...304...11M}) or within the solar neighbourhood
(e.g. Matteucci \& Greggio 1986~\cite{1986A&A...154..279M}). Within the current
simulations we are using a value of $A=0.1$.
With these ingredients and the mass dependent lifetime functions
$\tau(m)$, we can model the rate of SNIa as
\begin{equation}
R_{{\mathrm{SN\,Ia}}}(t) \, = \,
A\,\int\limits_{\displaystyle{M_{\mathrm{B,inf}}}}^{\displaystyle{M_{\mathrm{B,sup}}}}
\phi(m_{\mathrm B})
\int\limits_{\displaystyle{\mu_{\mathrm m}}}^{\displaystyle{\mu_{\mathrm M}}} 
f(\mu)\,
\psi(t-\tau_{m_2})\,{\mathrm d}\mu\,{\mathrm d}m_{\mathrm B}\,.
\end{equation}
where $M_{\mathrm{Bm}}$ and $M_{\mathrm{BM}}$ are the smallest and
largest values, respectively, that are allowed for the progenitor binary mass $m_{\mathrm B}$.
Then, the integral over $m_{\mathrm B}$ runs in the range between $M_{\mathrm{B,inf}}$ and
$M_{\mathrm{B,sup}}$, which represent the minimum and the maximum value of the
total mass of the binary system that is allowed to explode at the time
$t$. These values in general are functions of $M_{\mathrm{Bm}}$, $M_{\mathrm{BM}}$, and
$m_2(t)$, which in turn depend on the star formation history
$\Psi(t)$. In simulations, where the individual stellar particles are in commonly modelled
as an impulsive star formation event, $\psi(t)$ can therefore be approximated with a Dirac
$\delta$-function. The sum of all stellar particles and their individual formation time then
represent the complex star-formation history of the galaxies within the simulation.

\subsubsection{Supernova Type II, Low-, and Intermediate-mass stars} 

Computing the rates of SNII, low-mass stars (LMS), and
intermediate-mass stars (IMS) is conceptually simpler than calculating
the rates of SNIa, since they are purely driven by the lifetime
function $\tau(m)$ convolved with the star formation history $\psi(t)$
and multiplied  by the IMF $\phi(m=\tau^{-1}(t))$.  Since $\psi(t)$ is
a delta-function for the simple stellar populations used in
simulations, the SNII, LMS and IMS rates read 
\begin{equation}
R_{\mathrm{SNII|LMS|IMS}}(t)=\phi(m(t)) \times \left( -\frac{d\,m(t)}{d\, t}\right)
\end{equation}
where $m(t)$ is the mass of the star that dies at time $t$.  For AGB
rates, the above expression must be multiplied by a factor of $(1-A)$
if the mass $m(t)$ falls in the range of masses which is relevant for
the secondary stars of SNIa binary systems.

\subsubsection{The equations of chemical enrichment} 
In order to compute the total metal release from the simple stellar
population, we have to fold the above rates with the yields
$p^{\mathrm{SNIa|SNII|AGB}}_{Z_i}(m, Z)$ from SNIa, SNII and AGB stars
for a given element $i$ for stars born with initial metallicity $Z_i$,
and compute the evolution of the density $\rho_i(t)$ for each element
$i$ at each time $t$. As shown in Borgani et al.,
(2008)~\cite{Borgani2008}, this reads
\begin{eqnarray}
\dot{\rho}_i(t)=&-&\psi(t)Z_i(t) \\
&+&\int_{M_{BM}}^{M_{U}} \psi(t-\tau(m))p_{Z_i}^{\mathrm{SNII}}(m, Z)\varphi(m)\,dm \\
&+&A\int_{M_{Bm}}^{M_{BM}}\phi(m)\left[\int_{\mu_{\mathrm m}}^{\mu_{\mathrm M}}
f(\mu)\psi(t-\tau_{m_2})p_{Z_i}^{\mathrm{SNIa}}(m, Z)\,d\mu \right]\, dm \\
&+&(1-A)\int_{M_{Bm}}^{M_{BM}} \psi(t-\tau(m))p_{Z_i}^{\mathrm{AGB}}(m, Z)\varphi(m)\,dm \\
&+&\int_{M_L}^{M_{Bm}}
\psi(t-\tau(m))p_{Z_i}^{\mathrm{AGB}}(m, Z)\varphi(m)\,dm,\label{eq:stev}
\end{eqnarray}
where $M_{\mathrm L}$ and $M_{\mathrm U}$ are the minimum and maximum
mass of a star in the simple stellar population, respectively. Commonly adopted choices for these limiting masses are
$M_{\mathrm L}\simeq 0.1$~M$_\odot$ and $M_{\mathrm U}\simeq 100$~M$_\odot$.

In the above equation, the first line describes the locking of
metals in new born stars through the currently ongoing star formation
$\psi(t)$, which for the assumed sub-grid model case vanishes as $\psi(t)$ is a delta function.
For a comprehensive review of the analytic formalism we refer the reader to Greggio (2005)~\cite{Greggio2005}.

%%%%%%%%%%%%%%%%%%%%%%%%%%%%%%%%%%%%%%%%%%
\section{The Magneticum Simulations}
\begin{figure}[t]
\begin{center}
  \includegraphics[width=0.95\textwidth]{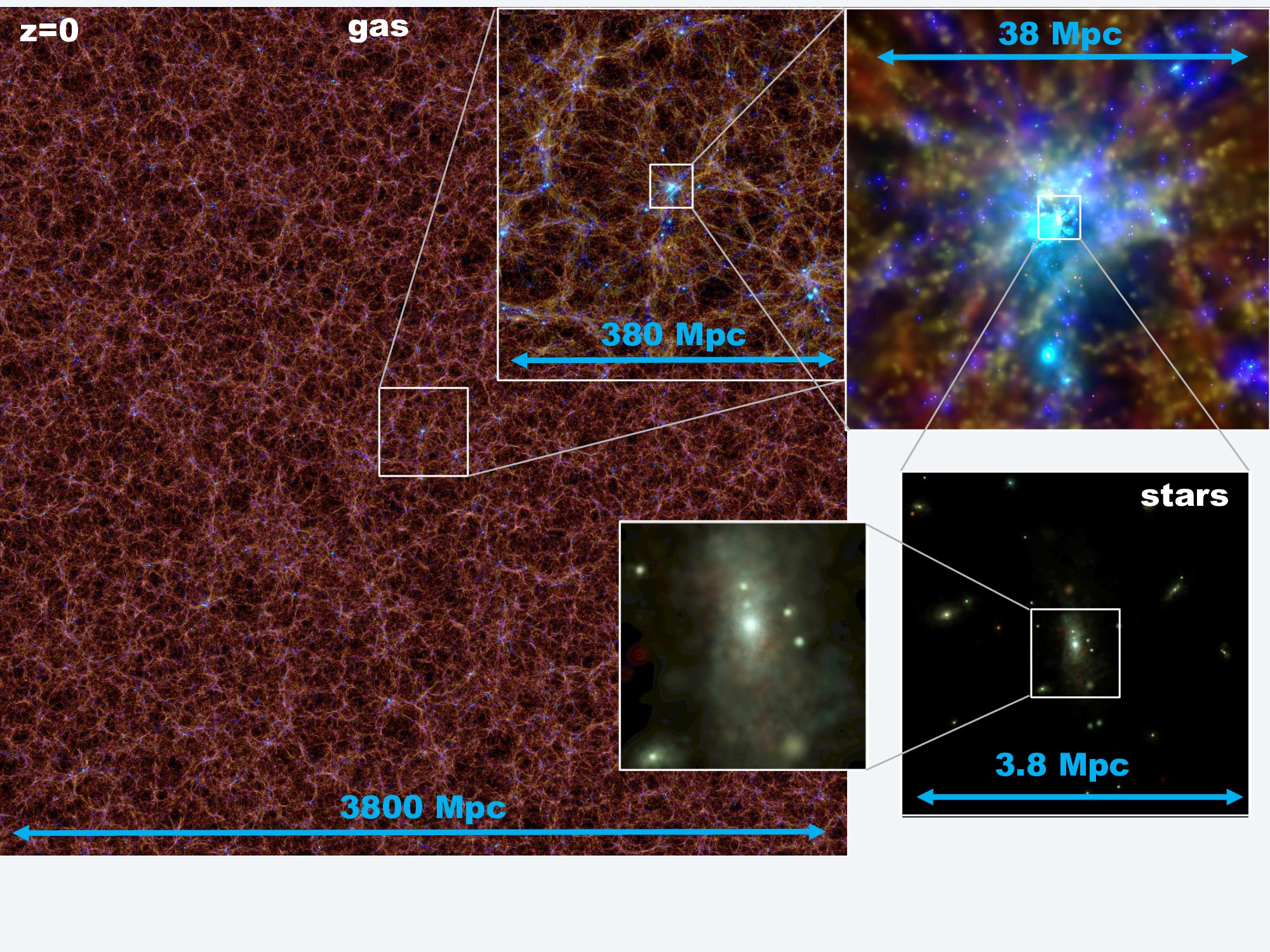}
  \caption{Visualization of the large scale distribution of the gas
    and stellar component within {\it Box0} (see Bocquet et al.,
    2016~\cite{2016MNRAS.456.2361B}) of the {\it Magneticum}
    simulation set at redshift $z=0$. The inlays show a consecutive
    zoom onto the most massive galaxy cluster, where the individual
    galaxies become visible.  }
   \label{fig1}
\end{center}
\end{figure}

\begin{table*}
\label{tab:Boxes}
\centering
\begin{tabular}{|c c c c c c c | c c | } 
\hline
Simulation   & Box0            & Box1            & Box2b          & Box2            & Box3           & Box4          & $m_\mathrm{star}$& $\epsilon_\mathrm{star}$ \\
\hline
Size [Mpc]   & 3820            & 1300            & 910            & 500             & 180            & 68            & [$M_\odot$]      & [kpc] \\
\hline
mr           & $2\times4536^3$ & $2\times1512^3$ & --              & $2\times594^3$  & $2\times216^3$ & $2\times81^3$ & $6.5\times10^8$           & 5\\
hr           & --              & --              & $2\times2880^3$ & $2\times1564^3$ & $2\times576^3$ & $2\times216^3$ & $3.5\times10^7$          & 2\\
uhr          & --              & --              & --              & --             & $2\times1536^3$ (z=2) & $2\times576^3$ & $1.9\times10^6$           & 0.7\\
\hline
\end{tabular}
\caption{Size and number of particles for the different simulations. The last two rows list the average mass and softening of the star particles for the different resolution levels.}
\label{tab1}
\end{table*}

The {\it Magneticum} simulation set covers a huge dynamical range,
from very large cosmological volumes as shown in Fig.~\ref{fig1},
which can be used for statistical studies of clusters and voids
(e.g. Bocquet et al.
2016~\cite{2016MNRAS.456.2361B}, Pollina et al. 2017~\cite{2017MNRAS.469..787P}), to very
high resolution simulations of smaller cosmological volumes which
allow a morphological classification and a detailed analysis of
galaxies and their properties (e.g. Teklu et al.
2015~\cite{Teklu2015} and Remus et al. 2017~\cite{Remus2017}).
Table \ref{tab1} lists the detailed properties like size and stellar mass-resolution
of the different simuations. These
simulations treat the metal-dependent radiative cooling, heating from
a uniform time-dependent ultraviolet background, star formation and
the chemo-energetic evolution of the stellar population as traced by
SNIa, SNII and AGB stars with the associated feedback processes and
stellar evolution details as described before.  They also include the
formation and evolution of super-massive black holes and the
associated quasar and radio-mode feedback processes. For a detailed
description of the simulation sample see Dolag et al. (in prep),
Hirschmann et al. (2014)~\cite{Hirschmann}, and Teklu et al.
(2015)~\cite{Teklu2015}.

%%%%%%%%%%%%%%%%%%%%%%%%%%%%%%%%%%%%%%%%%%
\section{Metallicities from Magneticum in Comparison to Observations}

As previously shown from re-simulations of massive galaxy clusters,
the observed radial profiles of Iron within the ICM can be well
reproduced in simulations: Biffi et al., (2017)~\cite{Biffi2017}
demonstrated that especially at high redshifts the implemented AGN
feedback is the main driver in enhancing the metal enrichment in the
ICM at large cluster-centric distances to the observed level. 

The {\it Magneticum} simulations now allow such investigations across
a much larger range of halo masses. For this study we use a large
simulation volume (Box2 hr) with enough resolution to resolve mean
properties of galaxies (and AGNs), resulting in the same resolution as
used in Biffi et al., (2017)~\cite{Biffi2017} and Hirschmann et al.,
(2014)~\cite{Hirschmann}.
\begin{figure}[t]
\begin{center}
  \includegraphics[width=0.5\textwidth]{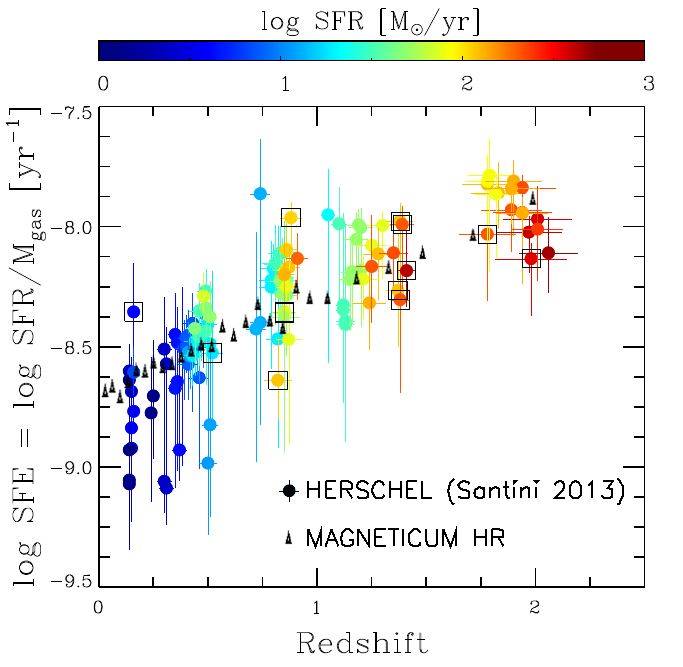}
  \caption{Redshift evolution of the SFE (coloured points with error
    bars) from Herschel observations presented by Santini et al.,
    (2014)~\cite{santini13} overlayed with the predicted median values
    at different redshifts from star forming galaxies in the {\it
      Magneticum} simulations (black triangles).  }
   \label{fig2}
\end{center}
\end{figure}
For galaxies, the star formation efficiency (SFE) is a measure of the
SFR per unit of gas mass. To evaluate whether the SF activity in the
simulations match the observational data is a key step to proceed
towards reproducing the observed mass-metallicity-relation (MZR). In
Fig.~\ref{fig2}, the evolution of the observed SFE up to a redshift of
$z\approx2$ is shown in comparison to the evolution of the SFE of star
forming galaxies in {\it Magneticum}. As can clearly be seen, the
simulations are in excellent agreement with the observed trends up to
$z\approx2$.

%%%%%%%%%%%%%%%%%%%%%%%%%%%%%%%%%%%%%%%%%%

\subsection{Galaxy Clusters: ICM Metallicities}
\begin{figure}[H]
\begin{center}
  \includegraphics[width=0.45\textwidth]{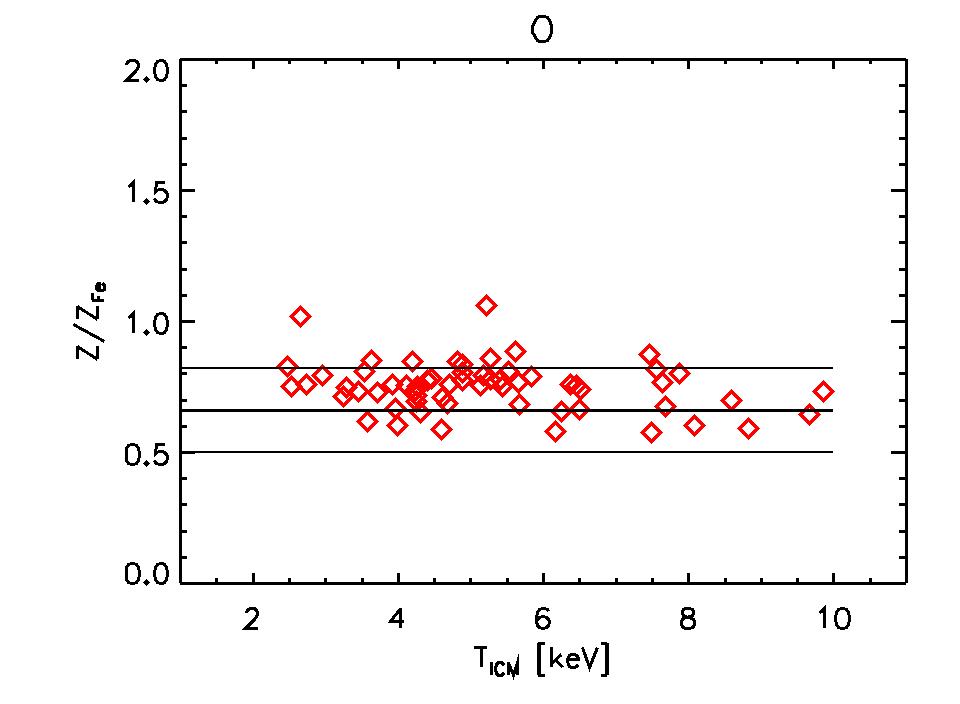}
  \includegraphics[width=0.45\textwidth]{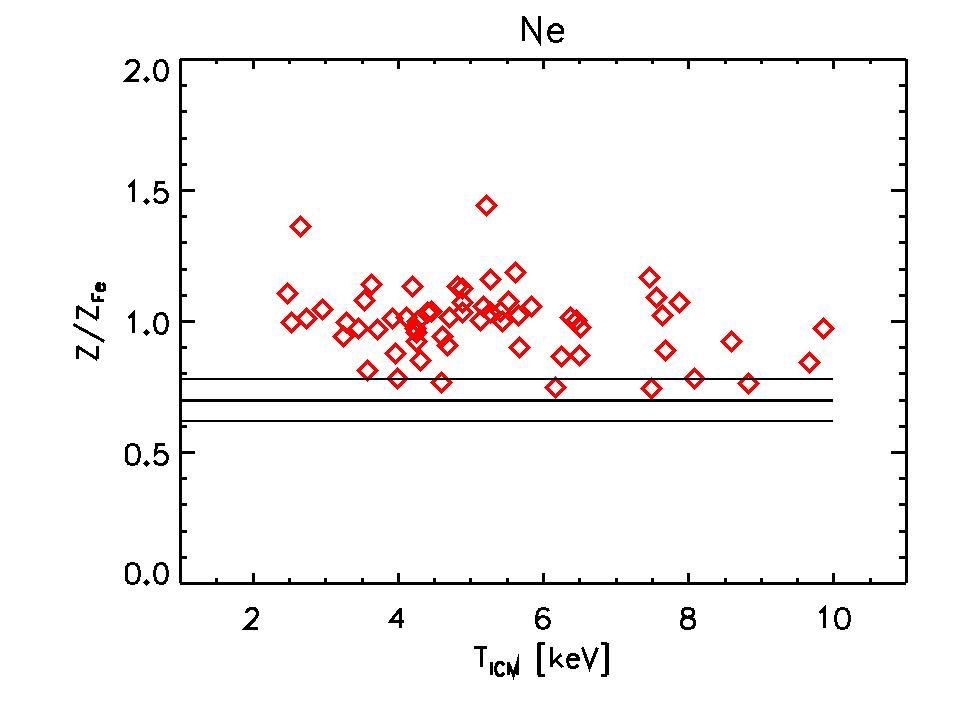}
  \includegraphics[width=0.45\textwidth]{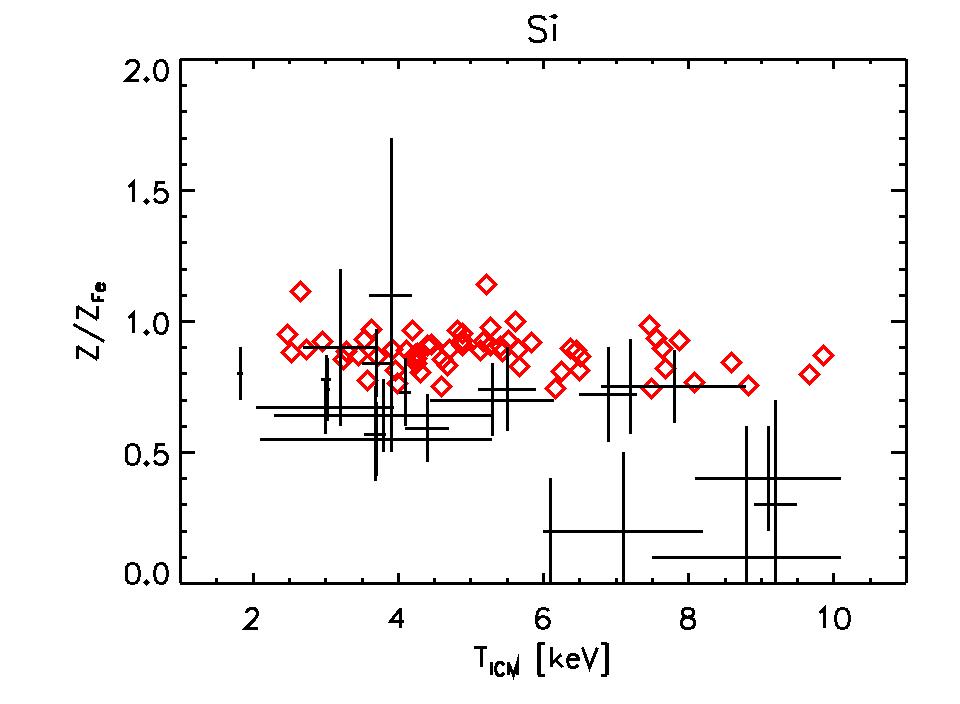}
  \includegraphics[width=0.45\textwidth]{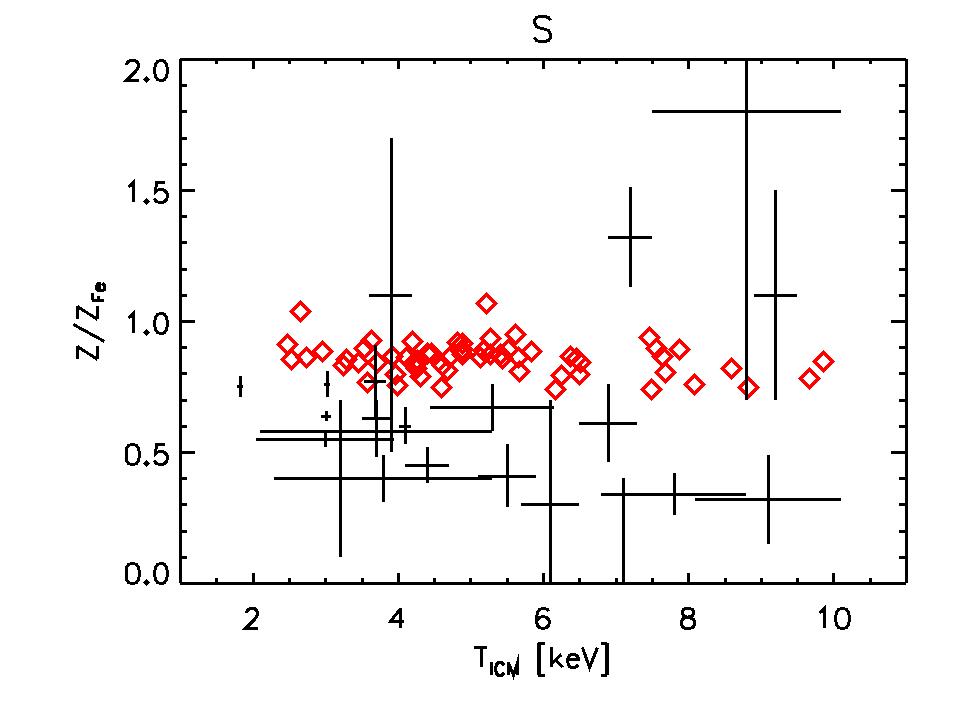}\\
  \includegraphics[width=0.45\textwidth]{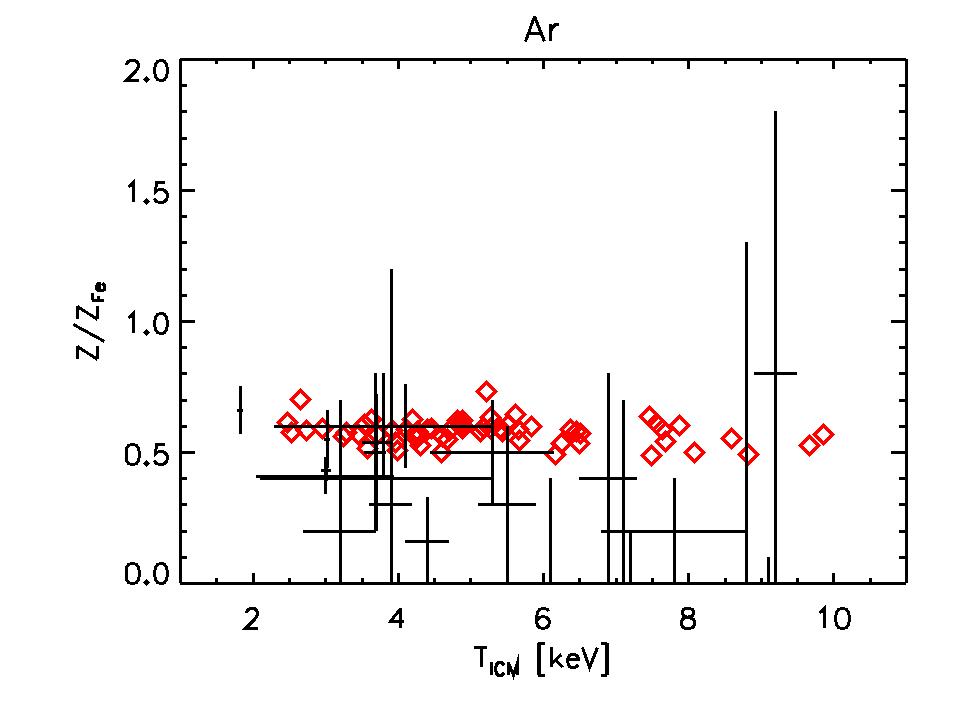}
  \includegraphics[width=0.45\textwidth]{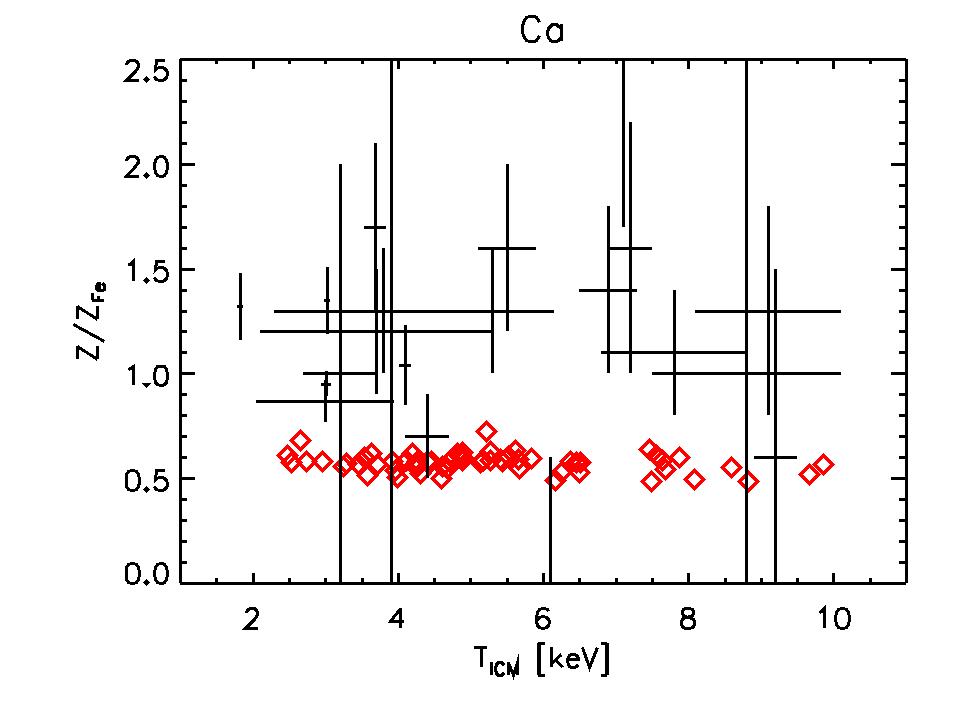}
  \includegraphics[width=0.45\textwidth]{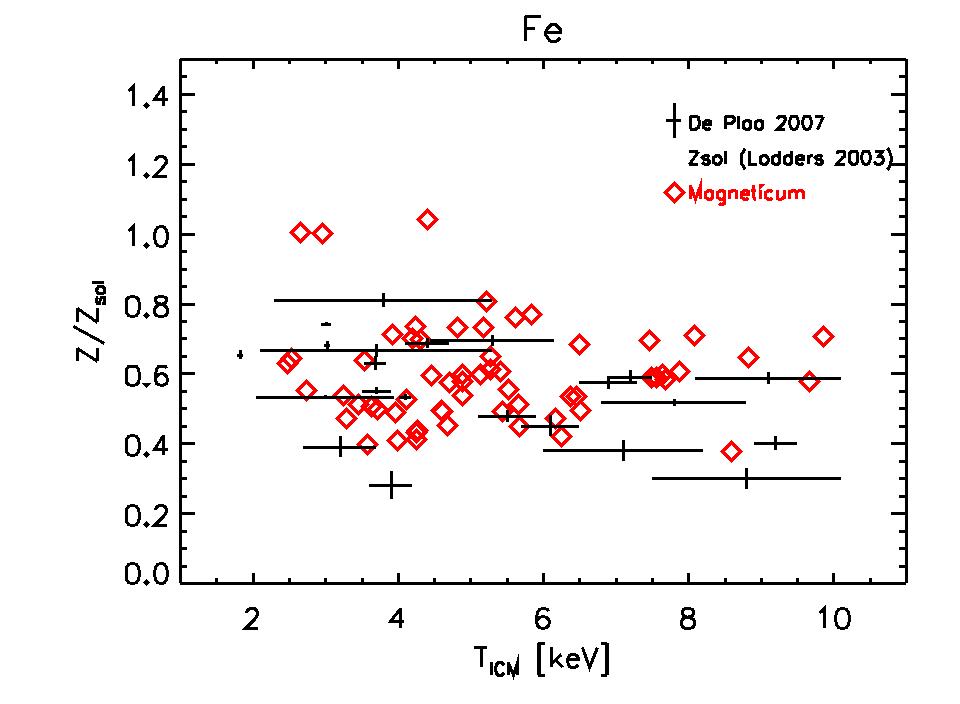}
  \includegraphics[width=0.45\textwidth]{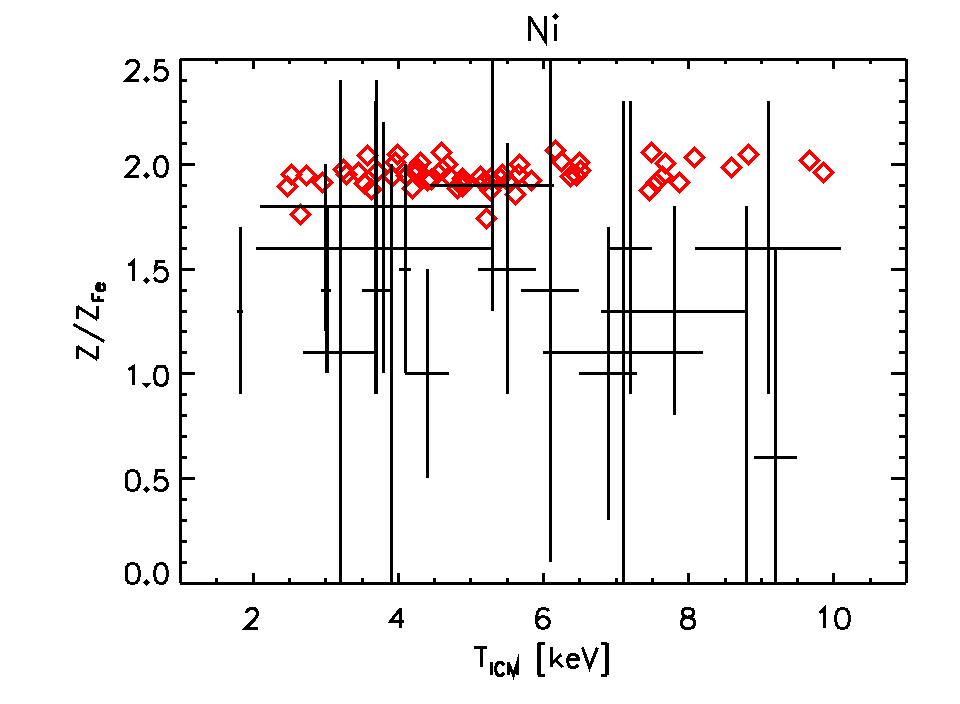}
  \caption{Comparison between the observed ICM metallicity within
    $R_{2500}$ (black symbols with error bars) and the metallicities
    {\bf of galaxy clusters} from the {\it Magneticum} simulations (red diamonds)
    {\bf as function of the} ICM temperature. Different panels show the ratios
    for different metal types, as indicated in the panels. From top
    left to bottom right the relative contribution of SNIa is expected
    to increases, while the relative contribution of SNII is expected
    to decrease. Observational data are taken from De Plaa et al.
    (2007)~\cite{2007A&A...465..345D}.  }
   \label{fig3}
\end{center}
\end{figure}

X-ray observations of the intra cluster medium allow for a detailed
study of the distribution of different metal species within the ICM
via their line emissions within the X-ray band.  Here, the composition
of the individual metal species allows to interpret their abundances
as an imprint of the relative contributions to the enrichment from
SNIa and SNII, hence keeping a record on when and where these metals
are injected into the ICM (e.g. De Plaa et al.,
2007~\cite{2007A&A...465..345D}). The lower left panel of Fig.~\ref{fig3}
shows a comparison of such observational data for various individual galaxy clusters with
\noindent
simulations, clearly demonstrating the
ability of the {\it Magneticum} simulations to reproduce the
correct absolute iron abundance within the ICM self-consistently.

Here the scatter of the absolute iron abundance within the simulated clusters
shows a similar spread (factor of two) as the observed values with a very mild
trend of increasing iron abundance for low mass (e.g. lower ICM temperature) clusters.
In addition the simulations also reproduce broadly the chemical
composition footprint of various elements species, as shown in the
various panels. This strongly
indicates that the simulations predict the correct ratio between the
contributions to the metal enrichment from SNIa and SNII and its
interplay with the AGN feedback. The simulations also predict
  typically less spread in the composition of the metals for the indicidual
  clusters than they show in their absolute iron metalicity.

%%%%%%%%%%%%%%%%%%%%%%%%%%%%%%%%%%%%%%%%%%
\subsection{Galaxies: Gas Metallicities}
\begin{figure}[t]
\begin{center}
  \includegraphics[width=0.49\textwidth]{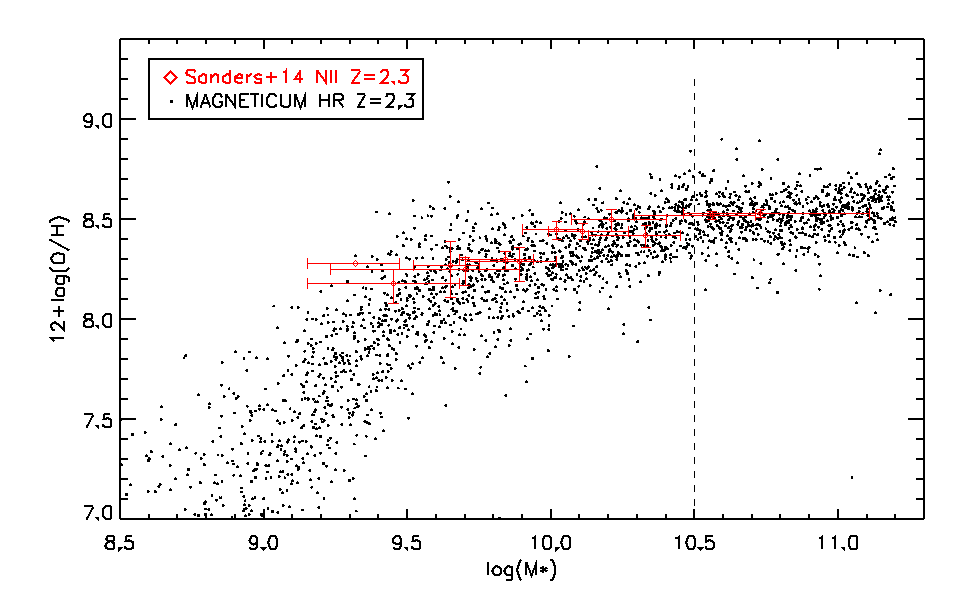}
  \includegraphics[width=0.49\textwidth]{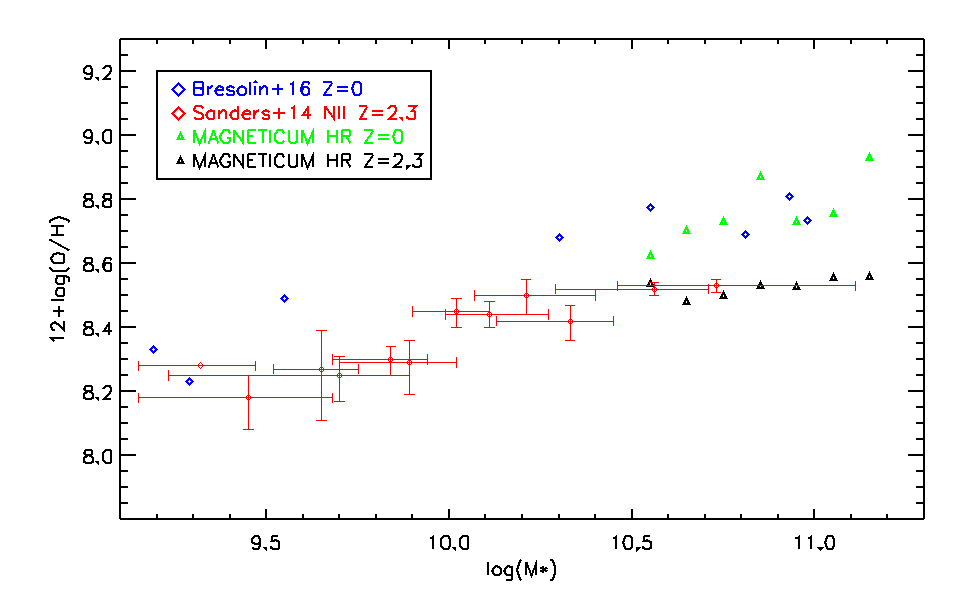}
  \caption{Gas-phase metallicity from {\it Magneticum} galaxies in
    comparison to observational data by Sanders et al.
    (2015)~\cite{sanders14} at redshift $z=2.3$ (red datapoints with
    error bars) and Bresolin et al. (2016)~\cite{bresolin16} at
    redshift $z=0$ (blue diamonds). \textit{Left panel:} Each black
    point represents a galaxy from the simulation. \textit{Right
      panel:} Median values for each mass bin at $z=2.3$ (black) and
    $z=0$ (green) are shown as triangles.  }
   \label{fig4}
\end{center}
\end{figure}

To estimate the gas phase metallicity of galaxies, it is important
consider that observationally the measurements are obtained only from
star forming regions. Therefore, it can be misleading to only
calculate the mean metallicity of all gas particles inside a simulated
galaxy. Thus, after selecting star forming galaxies (see Teklu et al.
2017~\cite{2017arXiv170206546T}) we can either calculate their mean
metallicity by averaging over all particles which are currently
star-forming, or alternatively calculate the mean metallicity of the
new-born stars.  We tested both methods and found that the latter
gives slightly better results due to the fact that it is difficult to
catch the metallicity of the gas phase in the moment of star formation
within simulations given the large timespan between the simulation
outputs, while the young stellar population freezes the record of the
metallicity of the gas from which it was formed. This leads to a MZR
which is in good agreement with observations, as shown in the left
panel of Fig.~\ref{fig4}, where we use the predicted Oxygen abundances
from our calculations to be consistent with observations (Sanders et
al. 2015~\cite{sanders14} at $z=2.3$ and Bresolin et al.
2016~\cite{bresolin16} at $z=0$).
Interestingly, even the overall evolution of the MZR is well captured
by the simulations and matches the observations, as demonstrated in
the right panel of Fig.~\ref{fig4}. However, when calculating
gas-phase metallicity gradients within the galaxies, the simulations
predict a steeper profile than the observations. Given that the
prediction of the mean metallicities is in good agreement with
observations this indicates that the simulations either still lack the
resolution to properly describe the mixing of the enriched gas within
galaxies, or that on these scales the diffusion of metals might have
to be modelled more explicitly. Nevertheless, we clearly showed that
including detailed modelling of the stellar population combined with
current AGN feedback models significantly improve the predicted ICM
and IGM metallicities and are needed to successfully reproduce various
aspects of the observations.

%%%%%%%%%%%%%%%%%%%%%%%%%%%%%%%%%%%%%%%%%%
\subsection{Galaxies: Stellar Metallicities}
In general, we assume that the mean metallicity of a stellar particle
in the simulation represents the mean stellar metallicity of the
stellar population represented by the stellar particle, neglecting any
self-enrichment within stars. To be consistent with the observations,
we based all calculations on the Iron abundance as predicted by the
simulations. For this part of the study we used a smaller cosmological
volume with a higher resolution, as this allows a classification of
the galaxies due to their morphological type (see Teklu et al.
2015~\cite{Teklu2015}) for more details on the classification and this
particular simulation). This higher resolution also allows for a more
detailed resolution of the metallicity gradients with radius for
individual galaxies.

\begin{figure}[t]
\begin{center}
  \includegraphics[width=0.49\textwidth]{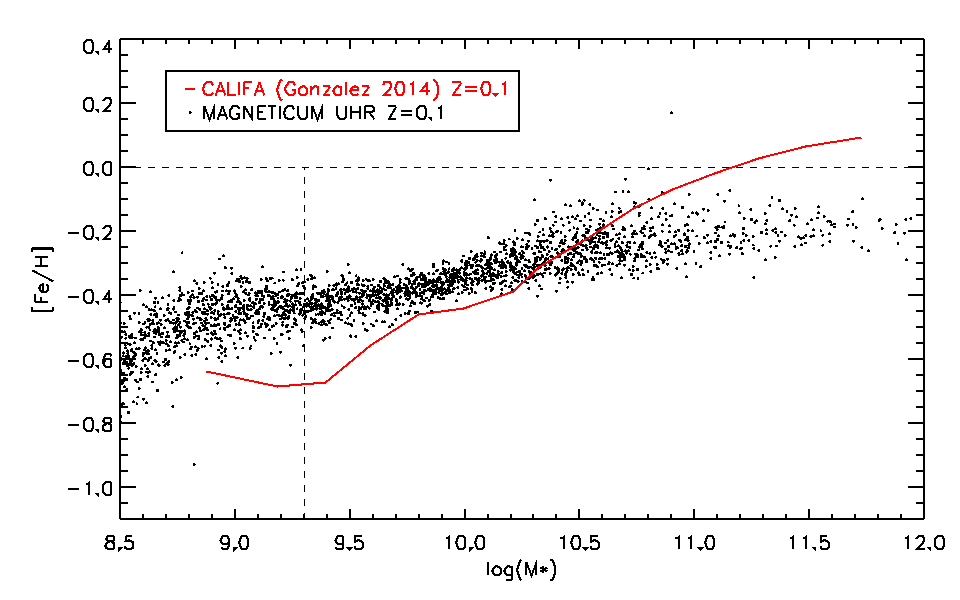}
  \includegraphics[width=0.49\textwidth]{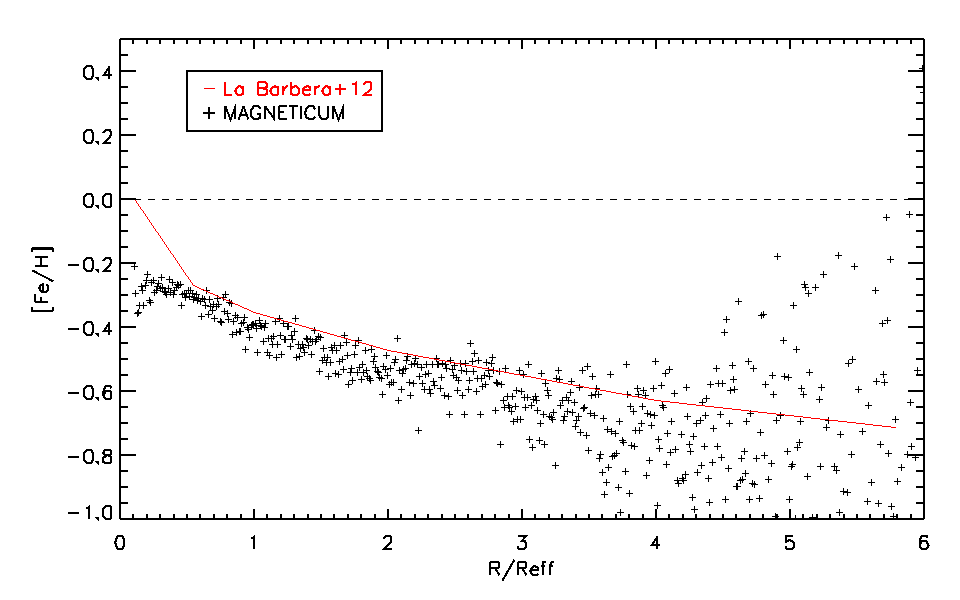}
  \caption{\textit{Left panel:} Stellar metallicity versus stellar
    mass obtained from the {\it Magneticum} galaxies (black dots) in
    comparison with observational data for galaxies independent of
    their morphology from Gonzalez Delgado et al.
    (2014)~\cite{gonzalez14} at redshift $z=0$.  \textit{Right panel:}
    Mean stellar radial metallicity gradient from 100 {\it Magneticum}
    galaxies in 100 radial bins normalized to the half-mass radius
    (black dots) compared to the observed profile from La Barbera et
    al. (2012)~\cite{laBarbera12}, red line.  }
  \label{fig5}
\end{center}
\end{figure}

Although the obtained mean metallicities of our galaxies are close to
observational results, the stellar MZR obtained from the simulations
is somewhat shallower that the one obtained from CALIFA observations
by Gonzalez Delgado et al. (2014)~\cite{gonzalez14}, as shown in the
left panel of Fig.~\ref{fig5}. This again indicates that the treatment
of the mixing between accreted, more pristine material from outside
the galaxy and the enriched gas within the galaxies is not fully
captured yet.  At this point it is unclear whether that will be
resolved by further enhancing the resolution of simulations or whether
explicit diffusion of metals has to be taken into account.

For a proper comparison of the radial metallicity gradients between
simulations and observational data, it is necessary to calculate the
effective radius $R_{\rm eff}$ for each galaxy. This is done by
selecting all stellar particles within ten percent of the virial
radius and then inferring the according half-mass radius, which we
associate to the effective radius $R_{\rm eff}$. This method has
already been used to demonstrate that the {\it Magneticum} simulations
successfully reproduce the mass-radius relation of galaxies for
different morphological types (e.g. Remus et al.
2015~\cite{2015IAUS..309..145R}) and at different redshifts
(e.g. Remus et al. 2017~\cite{Remus2017}).

Interestingly, the radial stellar metallicity gradients obtained from
the simulations very well match the CALIFA observations, as shown in
the right panel of Fig.~\ref{fig5}. Note also that the increasing
spread towards larger radius is an intrinsic, point-by-point spread
within the individual galaxies and agrees well with the behaviour
measured for individual galaxies by Pastorello et
al. (2014)~\cite{2014MNRAS.442.1003P}. However, while the simulations
successfully reproduce the observed Iron abundances and radial
gradients, they still produce a flat ratio of Oxygen (or Magnesium or
Silicium) over Iron, in contrast to the observations, and here the
sub-grid model clearly needs to be advanced. 

%%%%%%%%%%%%%%%%%%%%%%%%%%%%%%%%%%%%%%%%%%
\section{Discussion and Conclusion}
Metals are measured in all phases of the baryonic universe, starting
from the ICM in galaxy clusters down to gas and stars in individual
galaxies. As the radial distribution of different types of metals
encode a record of the star formation process over the whole time of a
galaxies' evolution, comparing the chemical footprint as seen in
simulations with observations reflects an excellent test on the
reliability of the numerical sub-grid models included in modern
state-of-the-art simulations of galaxy formation in a cosmological
context in reproducing the complex processed enrolled in galaxy
formation.

We demonstrated that the {\it Magneticum} simulations are able to
reproduce a large variety of observational findings over a large range
of halo masses, indicating that the star-formation and feedback
processes included start to be highly realistic. At galaxy cluster
scale the simulations show an excellent agreement with both, the absolute
value as well as the cluster by cluster variation of the the iron abundance
when comparing to x-ray observations and also reproduce broadly the chemical
composition of the ICM. On galaxy scales the tight
mass-metallicity-relations found for our simulated galaxies indicate
that there are no strong secondary parameters (like star formation
rates at fixed redshift) driving the scatter in these relations. The
remaining differences between the observed properties and the
simulation results indicate, however, that the incorporation of
physical processes like diffusion and mixing have to be improved
within the next generation of simulations to successfully reproduce
the detailed distribution of metals on individually resolved galaxy
scales.

%%%%%%%%%%%%%%%%%%%%%%%%%%%%%%%%%%%%%%%%%%
\acknowledgments{The Magneticum Pathfinder simulations were performed
  at the Leibniz-Rechenzentrum with CPU time assigned to the Projects
  ``pr86re'' and ``pr83li''.  This work was supported by the DFG
  Cluster of Excellence ``Origin and Structure of the Universe''.  We
  are especially grateful for the support by M. Petkova through the
  Computational Center for Particle and Astrophysics (C2PAP).}

%%%%%%%%%%%%%%%%%%%%%%%%%%%%%%%%%%%%%%%%%%
\authorcontributions{K.D. performed the simulation and wrote the
  paper; E.M. analysed the data; R.-S.R. contributed to the physical
  interpretation of the results, supervising E.M. during his master
  and to the writing of the paper.}

%%%%%%%%%%%%%%%%%%%%%%%%%%%%%%%%%%%%%%%%%%
\conflictsofinterest{The authors declare no conflict of interest.}

%=====================================
% References, variant B: external bibliography
%=====================================
\externalbibliography{yes}
\bibliography{ref_dolag}

\end{document}